\documentclass[journal]{IEEEtran}
\ifCLASSINFOpdf
  \usepackage[pdftex]{graphicx}
  \graphicspath{{../pdf/}{../jpeg/}}
  \DeclareGraphicsExtensions{.pdf,.jpeg,.png}
\else
  \usepackage[dvips]{graphicx}
  \graphicspath{{../eps/}}
  \DeclareGraphicsExtensions{.eps}
\fi
\usepackage{epsfig}
\usepackage{booktabs}
\usepackage{epstopdf}
\usepackage{amssymb}
\usepackage{mathtools}

\usepackage{algorithmic}

\usepackage{array}

\usepackage{mdwmath}
\usepackage{mdwtab}
\usepackage{bm}
\usepackage{mathrsfs}
\usepackage{epsfig}
\usepackage{subfigure}
\usepackage{algorithmic}
\usepackage{amsmath}
\usepackage{mathrsfs}

\begin{document}

\title{Different Power Adaption Methods on Fluctuating Two-Ray Fading Channels
\thanks{The authors are with the Computer, Electrical, and Mathematical Science and Engineering (CEMSE) Division, King Abdullah University of Science and Technology (KAUST), Thuwal, Makkah Province, Saudi Arabia
(email: \{hui.zhao; zhedong liu; slim.alouini\}@kaust.edu.sa).}
}

\author{Hui Zhao, Zhedong Liu, and Mohamed-Slim Alouini}

\maketitle

\begin{abstract}
In this letter, we consider a typical scenario where the transmitter employs different power adaption methods, including the optimal rate and power algorithm, optimal rate adaption, channel inversion and truncated channel inversion, to enhance the ergodic capacity (EC) with an average transmit power constraint  over  fluctuating two-way fading channels. In particular, we derive exact closed-form expressions for the EC under different power adaption methods, as well as  corresponding asymptotic formulas for the EC  valid in the high signal-to-noise ratio (SNR) region.  Finally, we compare the performance of  the EC under different power adaption methods, and this also validates the accuracy of our derived expressions for the exact and asymptotic EC.

\end{abstract}

\begin{IEEEkeywords}
 Asymptotic ergodic capacity, ergodic capacity,  fluctuating two-ray fading channel, power adaption.
\end{IEEEkeywords}

\section{Introduction}
Due to the exponential increase in aggregate traffic, millimeter-wave (mmWave) has been recently used to overcome the wireless spectrum shortage. Although some conventional fading channels, such as Rayleigh and Rician fading channels, has been verified to suit sometimes the mmWave radio communications, the fluctuation suffered by the received signal cannot always be modeled accurately by conventional fading models. In view of this issue, \cite{Goldsmith} introduced the fluctuating two-ray (FTR) fading model consisting of random phase plus a diffuse component, which is the natural generalization of the two-wave with diffuse (TWDP) model in \cite{Wang} where the specular components of the TWDP model are just constant amplitudes, and also can reduce to many conventional fading models, such as Rician and Nakagami-$m$ fading models (more special cases refer to the Table I in \cite{Goldsmith}). Subsequently, \cite{Zhang1,Zhang2} extended the work of \cite{Goldsmith} in terms of elementary functions and coefficients consisting of
fading parameters, where the parameter $m$ in FTR fading can be valued by an arbitrary positive real number, rather than only positive integers in \cite{Goldsmith}. However, there is no derivation of asymptotic ergodic capacity (AEC) and asymptotic ergodic secrecy capacity (AESC) in the high signal-to-noise ratio (SNR) region in \cite{Zhang1,Zhang2}, resulting in missing some insights and less efficiency in calculation of the ergodic capacity (EC) and ergodic secrecy capacity (ESC) in the high SNR region. Further, the authors in \cite{Zhang1,Zhang2} only consider the optimal rate adaption (ORA) case, where the transmit power is fixed over the whole transmission.

In practice, there is an average transmit power constraint at the transmitter, which will involve power adaption according to instantaneous channel state to enhance the EC. Optimal power and rate algorithm (OPRA) and channel inversion (CI) are two comment methods of power adaption \cite{Slim}-\cite{Pan}, where the performance of OPRA is much better than that of CI, because a large amount of the transmit power is required to compensate for the deep channel fading. To improve the performance of CI, \cite{Gold} proposed a kind of truncated CI (TCI), where the cutoff level can be selected to achieve a specified outage probability.

In this letter, we  derive exact and corresponding asymptotic closed-form expressions for the EC over FTR fading channels under different power adaption methods, namely OPRA, ORA, CI and TCI, and  compare the EC among them by simulation. 

\section{System Model}
The EC is defined as
$
\overline C  = \mathbb{E}\left\{ {\ln \left( {1 + {\gamma _d}} \right)} \right\} = \int_0^\infty  {\ln \left( {1 + {\gamma _d}} \right)} {f_{{\gamma _d}}}\left( {{\gamma _d}} \right)d{\gamma _d},
$
where $\gamma_d$ is the instantaneous SNR at the destination, and $f_{\gamma_d}(\cdot)$ is the probability density function (PDF) of $\gamma_d$. In the following sections, we will investigate several adaptive transmission methods to improve the EC.

The PDF and cumulative density function (CDF) of $\gamma_d$ over FTR fading channels are given by \cite{Zhang1}
\begin{align}
&{f_{{\gamma _d}}}\left( {{\gamma _d}} \right) = \frac{{m_d^{{m_d}}}}{{\Gamma \left( {{m_d}} \right)}}\sum\limits_{{j_d} = 0}^\infty  {\frac{{K_d^{{j_d}}{d_{{j_d}}}}}{{{j_d}!{j_d}!}}} \frac{{\gamma _d^{{j_d}}}\exp \left( { - \frac{{{\gamma _d}}}{{2\sigma _d^2}}} \right)}{{{{\left( {2\sigma _d^2} \right)}^{{j_d} + 1}}}},\\
&{F_{{\gamma _d}}}\left( {{\gamma _d}} \right) = \frac{{m_d^{{m_d}}}}{{\Gamma \left( {{m_d}} \right)}}\sum\limits_{{j_d} = 0}^\infty  {\frac{{K_d^{{j_d}}{d_{{j_d}}}}}{{{j_d}!{j_d}!}}\Upsilon \left( {{j_d} + 1,\frac{{{\gamma _d}}}{{2\sigma _d^2}}} \right)},
\end{align}
respectively, where $\Gamma(\cdot)$ and $\Upsilon \left( { \cdot , \cdot } \right)$ denote the Gamma function and lower incomplete Gamma function \cite{Gradshteyn}, respectively.  $K_d$ is the average power ratio of the dominant waves and remaining diffuse multipath,  $m_d$ is the parameter of Gamma distribution with unit mean,  $\sigma_d^2$ is the variance of the real (or imaginary) diffuse component, and
\begin{align}
{d_{{j_d}}} &\buildrel \Delta \over = \sum\limits_{k = 0}^{{j_d}} {j_d \choose k} {\left( {\frac{{{\Delta _d}}}{2}} \right)^k}\sum\limits_{l = 0}^k {k \choose l} \Gamma \left( {{j_d} + {m_d} + 2l - k} \right) \notag\\
&\cdot\exp \left( {\frac{{\pi \left( {2l - k} \right)i}}{2}} \right){\left( {{{\left( {{m_d} + {K_d}} \right)}^2} - {{\left( {{K_d}{\Delta _d}} \right)}^2}} \right)^{\frac{{ - \left( {{j_d} + {m_d}} \right)}}{2}}} \notag\\
&\cdot P_{{j_d} + {m_d} - 1}^{k - 2l}\left( {\frac{{{m_d} + {K_d}}}{{\sqrt {{{\left( {{m_d} + {K_d}} \right)}^2} - {{\left( {{K_d}{\Delta _d}} \right)}^2}} }}} \right),
\end{align}
in which $\Delta_d$, $i$, and $P(\cdot)$ denote a  ratio defined by (4) in \cite{Zhang1}, the imaginary unit, and the Legendre function of the first kind \cite{Gradshteyn}, respectively.
From ${F_{{\gamma _d}}}\left( \infty  \right) = 1$ and Eqs. (7) and (8) in \cite{Zhang1}, we can easily derive $\frac{{m_d^{{m_d}}}}{{\Gamma \left( {{m_d}} \right)}}\sum\nolimits_{{j_d} = 0}^\infty  {\frac{{K_d^{{j_d}}{d_{{j_d}}}}}{{{j_d}!}} = 1}$, and further rewrite $F_{\gamma_d}(\cdot)$ as
\begin{align}
{F_{{\gamma _d}}}\left( {{\gamma _d}} \right)=& 1 - \frac{{m_d^{{m_d}}}}{{\Gamma \left( {{m_d}} \right)}}\sum\limits_{{j_d} = 0}^\infty  {\frac{{K_d^{{j_d}}{d_{{j_d}}}}}{{{j_d}!}}} \notag\\
&\cdot\exp \left( { - \frac{{{\gamma _d}}}{{2\sigma _d^2}}} \right)\sum\limits_{n = 0}^{{j_d}} {\frac{1}{{n!}}{{\left( {\frac{{{\gamma _d}}}{{2\sigma _d^2}}} \right)}^n}}.
\end{align}

\section{Ergodic Capacity under OPRA}
\subsection{Exact EC under OPRA}
Our goal is to adjust the transmit power according to the instantaneous channel state to maximize the EC subject to a certain average transmit power ($\overline P_t$) by using OPRA, where the EC in the integral form is given by (7) in \cite{Slim}
\begin{align} \label{int_ESC}
&{\overline C_{opra}}= \int_{{\gamma _0}}^\infty  {\ln \left( {\frac{{{\gamma _d}}}{{{\gamma _0}}}} \right)} {f_{{\gamma _d}}}\left( {{\gamma _d}} \right)d{\gamma _d},
\end{align}
where $\gamma_0=\lambda \overline P_t$, in which $\lambda \ge 0$ is the corresponding Lagrangian multiplier.
Substituting the PDF of $\gamma_d$ into \eqref{int_ESC}, we can derive the EC under OPRA as
\begin{align}
{\overline C_{opra}}=& \frac{{m_d^{{m_d}}}}{{\Gamma \left( {{m_d}} \right)}}\sum\limits_{{j_d} = 0}^\infty  {\frac{{K_d^{{j_d}}{d_{{j_d}}}}}{{{j_d}!{j_d}!{{\left( {2\sigma _d^2} \right)}^{{j_d} + 1}}}}} \notag\\
&\cdot\underbrace {\int_{{\gamma _0}}^\infty  {\ln \left( {\frac{{{\gamma _d}}}{{{\gamma _0}}}} \right)} \gamma _d^{{j_d}}\exp \left( { - \frac{{{\gamma _d}}}{{2\sigma _d^2}}} \right)d{\gamma _d}}_{\mathcal{I}_1},
\end{align}
where
\begin{align}
{\mathcal{I}_1}  &= \int_0^\infty  {\ln \left( {\frac{{x + {\gamma _0}}}{{{\gamma _0}}}} \right)} {\left( {x + {\gamma _0}} \right)^{{j_d}}}\exp \left( { - \frac{{x + {\gamma _0}}}{{2\sigma _d^2}}} \right)dx \notag\\
 &= \sum\limits_{p = 0}^{{j_d}} {j_d \choose p} \gamma _0^{{j_d} - p}\exp \left( { - \frac{{{\gamma _0}}}{{2\sigma _d^2}}} \right)\notag\\
&\hspace{0.5cm} \cdot\int_0^\infty  {{x^p}\ln \left( {1 + \frac{x}{{{\gamma _0}}}} \right)} \exp \left( { - \frac{x}{{2\sigma _d^2}}} \right)dx.
\end{align}
By using the integral identity derived in Appendix B of \cite{Slim}, $\mathcal{I}_1$ can be easily solved in closed-form as
\begin{align}
{\mathcal{I}_1}= \sum\limits_{p = 0}^{{j_d}} {j_d \choose p} p!\sum\limits_{l = 1}^{p + 1} {{{\left( {2\sigma _d^2} \right)}^l}\gamma _0^{{j_d} + 1 - l}} \Gamma \left( { - p - 1 + l,\frac{{{\gamma _0}}}{{2\sigma _d^2}}} \right),
\end{align}
where $\Gamma(\cdot,\cdot)$ denotes the complementary upper incomplete Gamma function \cite{Slim}.

The corresponding constraint condition  for the transmit power can be written as \cite{Slim}
\begin{align}\label{condition}
&\int_{{\gamma _0}}^\infty  {\left( {\frac{1}{{{\gamma _0}}} - \frac{1}{{{\gamma _d}}}} \right)} {f_{{\gamma _d}}}\left( {{\gamma _d}} \right)d{\gamma _d} \notag\\
 &= \frac{1}{{{\gamma _0}}}\left[ {1 - {F_{{\gamma _d}}}\left( {{\gamma _0}} \right)} \right] - \int_{{\gamma _0}}^\infty  {\frac{1}{{{\gamma _d}}}} {f_{{\gamma _d}}}\left( {{\gamma _d}} \right)d{\gamma _d}=1.
\end{align}
Substituting the CDF and PDF of $\gamma_d$ over FTR fading channels into \eqref{condition}, we can derive \eqref{condition1} shown on the top of next page.  In view of the definition of upper incomplete Gamma function $\Gamma(\cdot,\cdot)$, we finally have the constraint condition as
\begin{figure*}
\begin{align}\label{condition1}
 \frac{1}{{{\gamma _0}}}\frac{{m_d^{{m_d}}}}{{\Gamma \left( {{m_d}} \right)}}\sum\limits_{{j_d} = 0}^\infty  {\frac{{K_d^{{j_d}}{d_{{j_d}}}}}{{{j_d}!}}} \exp \left( { - \frac{{{\gamma _0}}}{{2\sigma _d^2}}} \right)\sum\limits_{n = 0}^{{j_d}} {\frac{1}{{n!}}{{\left( {\frac{{{\gamma _0}}}{{2\sigma _d^2}}} \right)}^n}} - \frac{{m_d^{{m_d}}}}{{\Gamma \left( {{m_d}} \right)}}\sum\limits_{{j_d} = 0}^\infty  {\frac{{K_d^{{j_d}}{d_{{j_d}}}}}{{{j_d}!{j_d}!{{\left( {2\sigma _d^2} \right)}^{{j_d} + 1}}}}} \int_{{\gamma _0}}^\infty  {\gamma _d^{{j_d} - 1}} \exp \left( { - \frac{{{\gamma _d}}}{{2\sigma _d^2}}} \right)d{\gamma _d}=1.
\end{align}
\rule{18cm}{0.01cm}
\end{figure*}
\begin{align}\label{condition_f}
&\frac{{m_d^{{m_d}}}}{{\Gamma \left( {{m_d}} \right)}}\sum\limits_{{j_d} = 0}^\infty  {\frac{{K_d^{{j_d}}{d_{{j_d}}}}}{{{j_d}!{j_d}!}}} \notag\\
& \cdot\left\{ {\frac{1}{{{\gamma _0}}}\Gamma \left( {{j_d} + 1,\frac{{{\gamma _0}}}{{2\sigma _d^2}}} \right) - \frac{1}{{2\sigma _d^2}}\Gamma \left( {{j_d},\frac{{{\gamma _0}}}{{2\sigma _d^2}}} \right)} \right\} = 1.
\end{align}
Let
\begin{align}\label{condition_int}
f\left( {{\gamma _0}} \right) = \frac{1}{{{\gamma _0}}}\left[ {1 - {F_{{\gamma _d}}}\left( {{\gamma _0}} \right)} \right] - \int_{{\gamma _0}}^\infty  {\frac{{{f_{{\gamma _d}}}\left( {{\gamma _d}} \right)}}{{{\gamma _d}}}} d{\gamma _d}.
\end{align}
Differentiating $f(\gamma_0)$ with respect to $\gamma_0$ by using Leibniz Rule, we have
\begin{align}
\frac{{\partial f\left( {{\gamma _0}} \right)}}{{\partial {\gamma _0}}} =  - \frac{{1 - {F_{{\gamma _d}}}\left( {{\gamma _0}} \right)}}{{\gamma _0^2}}.
\end{align}
For ${\gamma _0} > 0$, $\frac{\partial }{{\partial {\gamma _0}}}f\left( {{\gamma _0}} \right) < 0$. Thus, $f\left( {{\gamma _0}} \right)$ is monotonically decreasing over ${\gamma _0} \in \left[ {0,\infty } \right)$. When $\gamma_0 \to \infty$,  \eqref{condition_int} will converge to 0, and when $\gamma_0 \to 0$, it will go to infinity. To summarize, there exists unique $\gamma_0$ satisfying the identity of  \eqref{condition_f}.  When $2\sigma _d^2 \to \infty$, we have \eqref{gamma0}, shown on the top of next page. Our numerical results show that $\gamma_0$ increases as $\overline \gamma_D$ increases, and as such $\gamma_0$ will lie in the interval $[0,1]$.
\begin{figure*}
\begin{align}\label{gamma0}
\frac{{m_d^{{m_d}}}}{{\Gamma \left( {{m_d}} \right)}}\sum\limits_{{j_d} = 0}^\infty  {\frac{{K_d^{{j_d}}{d_{{j_d}}}}}{{{j_d}!{j_d}!}}} \left\{ {\frac{1}{{{\gamma _0}}}\Gamma \left( {{j_d} + 1,\frac{{{\gamma _0}}}{{2\sigma _d^2}}} \right) - \frac{1}{{2\sigma _d^2}}\Gamma \left( {{j_d},\frac{{{\gamma _0}}}{{2\sigma _d^2}}} \right)} \right\}
 = \frac{1}{{{\gamma _0}}}\underbrace {\frac{{m_d^{{m_d}}}}{{\Gamma \left( {{m_d}} \right)}}\sum\limits_{{j_d} = 0}^\infty  {\frac{{K_d^{{j_d}}{d_{{j_d}}}}}{{{j_d}!}}} }_{ = 1} = 1 \Rightarrow {\gamma _0} = 1.
\end{align}
\rule{18cm}{0.01cm}
\end{figure*}
\subsection{Asymptotic EC under OPRA}
By using the limit identity $
\mathop {\lim }\limits_{x \to 0} \frac{{\Gamma \left( {s,x} \right)}}{{{x^s}}} =  - \frac{1}{s}$ for $\text{Re}(s)<0$ and $\mathop {\lim }\limits_{x \to 0} \Gamma \left( {0,x} \right) =  - \ln x + \psi \left( 1 \right)$, we have
\begin{align}
\mathop {\lim }\limits_{{1 \mathord{\left/
 {\vphantom {1 {\left( {2\sigma _d^2} \right)}}} \right.
 \kern-\nulldelimiterspace} {\left( {2\sigma _d^2} \right)}} \to 0} \frac{{\Gamma \left( { - p + l - 1,{1 \mathord{\left/
 {\vphantom {1 {\left( {2\sigma _d^2} \right)}}} \right.
 \kern-\nulldelimiterspace} {\left( {2\sigma _d^2} \right)}}} \right)}}{{{{\left( {{1 \mathord{\left/
 {\vphantom {1 {\left( {2\sigma _d^2} \right)}}} \right.
 \kern-\nulldelimiterspace} {\left( {2\sigma _d^2} \right)}}} \right)}^{ - p + l - 1}}}} = \frac{1}{{p + 1 - l}},
\end{align}
and
\begin{align}
&\mathop {\lim }\limits_{{1 \mathord{\left/
 {\vphantom {1 {\left( {2\sigma _d^2} \right)}}} \right.
 \kern-\nulldelimiterspace} {\left( {2\sigma _d^2} \right)}} \to 0} {\left( {\frac{{{\gamma _0}}}{{2\sigma _d^2}}} \right)^{{j_d} - p}}\left( {\sum\limits_{l = 1}^p {\frac{1}{{p + 1 - l}}}  + \Gamma \left( {0,\frac{{{\gamma _0}}}{{2\sigma _d^2}}} \right)} \right) \notag\\
&=\begin{cases}
0, &{p < {j_d}}; \\
{\psi \left( {p + 1} \right) - \ln \left( {\frac{{{\gamma _0}}}{{2\sigma _d^2}}} \right)}, &{p = {j_d}},
\end{cases}
\end{align}
where $\psi(\cdot)$ denotes the digamma function \cite{Gradshteyn}. In view of this limit relationship, the AEC under OPRA can be derived by
\begin{align} \label{C_inf}
\overline C_{opra}^\infty= \ln \left( {2\sigma _d^2} \right) - \ln {\gamma _0} + \frac{{m_d^{{m_d}}}}{{\Gamma \left( {{m_d}} \right)}}\sum\limits_{{j_d} = 0}^\infty  {\frac{{K_d^{{j_d}}{d_{{j_d}}}}}{{{j_d}!}}} \psi \left( {{j_d} + 1} \right).
\end{align}
Finally, $\overline C_{opra}^\infty $ can be further rewritten by using the relationship between $\sigma_d$ and $\overline \gamma_D$, i.e., ${\overline \gamma  _D} = \frac{{{E_b}}}{{{N_0}}}2\sigma _d^2\left( {1 + {K_d}} \right)r_d^{ - {\eta _d}}$, given by (3) in \cite{Zhang2}, where $E_b$, $N_0$, ${\eta _d}$ and $r_d$  are the energy per bit, noise power, path-loss exponent, and distance  between the transmitter and destination, respectively,
\begin{align}
\overline C_{opra}^\infty&= \ln {\overline \gamma  _D} - \ln \left( {\frac{{{E_b}\left( {1 + {K_d}} \right)r_d^{ - {\eta _d}}}}{{{N_0}}}} \right) \notag\\
&\hspace{0.5cm}- \ln {\gamma _0} + \frac{{m_d^{{m_d}}}}{{\Gamma \left( {{m_d}} \right)}}\sum\limits_{{j_d} = 0}^\infty  {\frac{{K_d^{{j_d}}{d_{{j_d}}}}}{{{j_d}!}}} \psi \left( {{j_d} + 1} \right).
\end{align}
We can easily see that the slope of EC in high SNRs is unity with respect to $\ln\overline \gamma_D$, and the power offset of EC in high SNRs  is independent of $\overline \gamma_D$, because when $\overline \gamma_D \to \infty$, $\gamma_0 \to 1$, which means that $\ln \gamma_0 \to 0$, and therefore, there is no impact of $\overline \gamma_D$ on the power offset in high SNRs.

\section{Ergodic Capacity under ORA}
\subsection{Exact EC under ORA}
In the ORA case, the transmitter cannot employ the channel state to adjust its transmit power instantaneously, and just uses a  constant power, i.e., average transmit power, to transmit signal to the destination.  From the Lemma 2 in \cite{Zhang1}, the EC under ORA is
\begin{align}
{\overline C_{ora} } =& \frac{{m_d^{{m_d}}}}{{\Gamma \left( {{m_d}} \right)}}\sum\limits_{{j_d} = 0}^\infty  {\frac{{K_d^{{j_d}}{d_{{j_d}}}}}{{{j_d}!}}} \exp \left( {\frac{1}{{2\sigma _d^2}}} \right)\notag\\
&\sum\limits_{l = 1}^{{j_d} + 1} {{{\left( {2\sigma _d^2} \right)}^{ - {j_d} - 1 + l}}} \Gamma \left( { - j_d - 1 + l,\frac{1}{{2\sigma _d^2}}} \right).
\end{align}
\subsection{Asymptotic EC under ORA}
We can  use $
\mathop {\lim }\limits_{x \to 0} \frac{{\Gamma \left( {s,x} \right)}}{{{x^s}}} =  - \frac{1}{s}$ for $\text{Re}(s)<0$ and $\mathop {\lim }\limits_{x \to 0} \Gamma \left( {0,x} \right) =  - \ln x + \psi \left( 1 \right)$ to derive the AEC under ORA in high SNR
\begin{align}\label{C_orcp}
{\overline C_{ora}^\infty } &= \frac{{m_d^{{m_d}}}}{{\Gamma \left( {{m_d}} \right)}}\sum\limits_{{j_d} = 0}^\infty  {\frac{{K_d^{{j_d}}{d_{{j_d}}}}}{{{j_d}!}}} \left( {\psi \left( {{j_d} + 1} \right) + \ln \left( {2\sigma _d^2} \right)} \right) \notag\\
 &= \ln \left( {2\sigma _d^2} \right) + \frac{{m_d^{{m_d}}}}{{\Gamma \left( {{m_d}} \right)}}\sum\limits_{{j_d} = 0}^\infty  {\frac{{K_d^{{j_d}}{d_{{j_d}}}}}{{{j_d}!}}\psi \left( {{j_d} + 1} \right)}.
\end{align}
This result is the same as \eqref{C_inf}, because $\gamma_0 \to 1$ as $\overline \gamma_D \to \infty$ in \eqref{C_inf}.

Another way to derive the AEC is to use the moments of $\gamma_d$, where the $n$th moment of $\gamma_d$ is given by
\begin{align}
\mathbb{E}\left\{ {\gamma _d^n} \right\} &= \int_0^\infty  {{x^n}} {f_{{\gamma _d}}}\left( x \right)dx \notag\\
&= \frac{{m_d^{{m_d}}}}{{\Gamma \left( {{m_d}} \right)}}\sum\limits_{{j_d} = 0}^\infty  {\frac{{K_d^{{j_d}}{d_{{j_d}}}}}{{{j_d}!{j_d}!}}} \Gamma \left( {n + {j_d} + 1} \right)\left( {2\sigma _d^2} \right)^n,
\end{align}
The first derivative of the $n$th moment with respect to $n$ is
\begin{align}
\frac{{\partial \mathbb{E} \left\{ {\gamma _d^n} \right\}}}{{\partial n}} =& \frac{{m_d^{{m_d}}}}{{\Gamma \left( {{m_d}} \right)}}\sum\limits_{{j_d} = 0}^\infty  {\frac{{K_d^{{j_d}}{d_{{j_d}}}}}{{{j_d}!{j_d}!}}} {\left( {2\sigma _d^2} \right)^n}\Gamma \left( {n + {j_d} + 1} \right) \notag\\
& \cdot \left[ {\ln \left( {2\sigma _d^2} \right) + \psi \left( {{j_d} + n + 1} \right)} \right].
\end{align}
In considering Eq. (20) in \cite{Slim2}, the AEC can be derived by
\begin{align}
\overline C _{ora}^\infty  &= {\left. {\frac{{\partial \mathbb{E}\left\{ {\gamma _d^n} \right\}}}{{\partial n}}} \right|_{n = 0}} \notag\\
&= \ln \left( {2\sigma _d^2} \right) + \frac{{m_d^{{m_d}}}}{{\Gamma \left( {{m_d}} \right)}}\sum\limits_{{j_d} = 0}^\infty  {\frac{{K_d^{{j_d}}{d_{{j_d}}}}}{{{j_d}!}}} \psi \left( {{j_d} + 1} \right),
\end{align}
which is the same as \eqref{C_orcp}.

\section{Ergodic Capacity under Channel Inversion}
\subsection{Exact EC under CI}
In the CI case, the transmit power is adjusted according to the channel state to maintain a constant SNR at the receiver, and the corresponding EC is given by (46) in \cite{Slim}
\begin{align}
{\overline C _{ci}} = \ln \left( {1 + \frac{1}{{\mathbb{E}\left\{ {{1 \mathord{\left/
 {\vphantom {1 {{\gamma _d}}}} \right.
 \kern-\nulldelimiterspace} {{\gamma _d}}}} \right\}}}} \right),
\end{align}
where
\begin{align}\label{CI}
&\mathbb{E}\left\{ {\frac{1}{{{\gamma _d}}}} \right\} = \frac{{m_d^{{m_d}}}}{{\Gamma \left( {{m_d}} \right)}}\sum\limits_{{j_d} = 0}^\infty  {\frac{{K_d^{{j_d}}{d_{{j_d}}}}}{{{j_d}!{j_d}!}}} \frac{{\Gamma \left( {{j_d}} \right)}}{{\left( {2\sigma _d^2} \right)}}.
\end{align}
When $d_{0}$ is not equal to zero or $K_d$ is not infinity, $\mathbb{E}\left\{ {\frac{1}{{{\gamma _d}}}} \right\}$ will become infinity, because $\Gamma(j_d) \to \infty$ for $j_d=0$, and thus EC under CI will be zero.
\subsection{Exact EC under TCI}
We consider the TCI case where a cutoff level $\gamma_0$ is selected to achieve a specified outage probability. In this case, the EC is given by Eq. (12) in \cite{Gold}
\begin{align}
{{\overline C}_{tci}} = \ln \left( {1 + \frac{1}{{\int_{{\gamma _0}}^\infty  {{1 \mathord{\left/
 {\vphantom {1 x}} \right.
 \kern-\nulldelimiterspace} x}{f_{{\gamma _d}}}\left( x \right)dx} }}} \right){\overline F _{{\gamma _d}}}\left( {{\gamma _0}} \right),
\end{align}
where $\overline F_{\gamma_d}(\cdot)$ is the complementary CDF of $\gamma_d$, and
\begin{align}
&\int_{{\gamma _0}}^\infty  {\frac{1}{x}} {f_{{\gamma _d}}}\left( x \right)dx
= \frac{{m_d^{{m_d}}}}{{\Gamma \left( {{m_d}} \right)}}\sum\limits_{{j_d} = 0}^\infty  {\frac{{K_d^{{j_d}}{d_{{j_d}}}}}{{{j_d}!{j_d}!}}\frac{1}{{\left( {2\sigma _d^2} \right)}}} \Gamma \left( {{j_d},\frac{{{\gamma _0}}}{{2\sigma _d^2}}} \right).
\end{align}
The EC under TCI  over FTR fading channel is
\begin{align}
&\overline C _{tci}  = \frac{{m_d^{{m_d}}}}{{\Gamma \left( {{m_d}} \right)}}\sum\limits_{{j_d} = 0}^\infty  {\frac{{K_d^{{j_d}}{d_{{j_d}}}}}{{{j_d}!{j_d}!}}} \Gamma \left( {{j_d} + 1,\frac{{{\gamma _0}}}{{2\sigma _d^2}}} \right) \notag\\
&\cdot\ln \left( {1 + {{\left( {\frac{{m_d^{{m_d}}}}{{\Gamma \left( {{m_d}} \right)}}\sum\limits_{{j_d} = 0}^\infty  {\frac{{K_d^{{j_d}}{d_{{j_d}}}}}{{{j_d}!{j_d}!}}\frac{1}{{\left( {2\sigma _d^2} \right)}}} \Gamma \left( {{j_d},\frac{{{\gamma _0}}}{{2\sigma _d^2}}} \right)} \right)}^{ - 1}}} \right).
\end{align}
\subsection{Asymptotic EC under TCI}
For $2\sigma_d^2 \to \infty$, it is easy to see that $\mathop {\lim }\limits_{2\sigma _d^2 \to \infty } {\overline F _{{\gamma _d}}}\left( {{\gamma _d}} \right) \to 1$. Then,  we consider the limit identity
\begin{align}
\mathop {\lim }\limits_{2\sigma _d^2 \to \infty } \frac{{\Gamma \left( {{j_d},\frac{{{\gamma _0}}}{{2\sigma _d^2}}} \right)}}{{\left( {2\sigma _d^2} \right)}} =
\begin{cases}
{\frac{{\Gamma \left( {{j_d}} \right)}}{{ {2\sigma _d^2} }}}, &{{j_d} > 0};\\
{\frac{{ - \ln \left( {{\gamma _0}} \right) + \ln \left( {2\sigma _d^2} \right) + \psi \left( 1 \right)}}{{2\sigma _d^2}}}, &{{j_d} = 0},
\end{cases}
\end{align}
where we truncate the Taylor expansion for ${{\Gamma \left( {{j_d},\frac{{{\gamma _0}}}{{2\sigma _d^2}}} \right)} \mathord{\left/
 {\vphantom {{\Gamma \left( {{j_d},\frac{{{\gamma _0}}}{{2\sigma _d^2}}} \right)} {\left( {2\sigma _d^2} \right)}}} \right.
 \kern-\nulldelimiterspace} {\left( {2\sigma _d^2} \right)}}$ at $2\sigma_d^2=\infty$ up to the first order term.
Let $\xi  =  - \ln \left( {{\gamma _0}} \right) + \ln \left( {2\sigma _d^2} \right) + \psi \left( 1 \right)$, and finally, the AEC under TCI is given by
\begin{align}\label{C_TCI}
&\overline C _{tci}^\infty  = \ln \left( {\frac{{2\sigma _d^2\Gamma \left( {{m_d}} \right)}}{{m_d^{{m_d}}\left( {{d_0}\xi  + \sum\nolimits_{{j_d} = 1}^\infty  {\frac{{K_d^{{j_d}}{d_{{j_d}}}\Gamma \left( {{j_d}} \right)}}{{{j_d}!{j_d}!}}} } \right)}}} \right) \notag\\
 &= \ln \left( {2\sigma _d^2} \right) + \ln \left( {\frac{{\Gamma \left( {{m_d}} \right)}}{{m_d^{{m_d}}}}} \right)- \ln \left( {{d_0}\xi  + \sum\limits_{{j_d} = 1}^\infty  {\frac{{K_d^{{j_d}}{d_{{j_d}}}\Gamma \left( {{j_d}} \right)}}{{{j_d}!{j_d}!}}} } \right).
\end{align}
From \eqref{C_TCI}, we can see that the EC under TCI is not a line function with respect to $\ln(2\sigma_d^2)$ or $\ln(\overline \gamma_D)$ in the high SNR region. However, the slope of \eqref{C_TCI} with respect to $\ln(2\sigma_d^2)$ changes very slowly in the high SNR region, which can be shown in Fig. 1 in the simulation section.

\section{Simulation}
In this section, we use Monte-Carlo simulation to validate our derived closed-form expressions for the EC and AEC in OPRA, ORA, CI, and TCI cases. In calculation of the infinite summation terms in the PDF and CDF of $\gamma_d$ over FTR fading channels, we can truncate the infinite summation terms into finite terms, where the resulting truncation error can be evaluated by (6) shown on the II-B subsection in \cite{Zhang2}.

As shown in Fig. 1, we can easily see that the EC increases with $\overline \gamma_D$ increasing, because of the improved channel state between the transmitter and receiver. It is also obvious that the EC is in decline as $m_d$ decreases, due to heavier fading, which is reflected by the larger power offset (intercept on horizontal axis). Moreover, the performance of EC under OPRA is better than that under ORA, because the transmitter adjusts its transmit power according to the instantaneous channel state in the OPRA case, rather than just using the fix power (average power) to transmit signal in another case. It is also worth to note that  the EC in OPRA and ORA cases converges  in the high SNR region, due to the fact that $\gamma_0 \to 1$ as $\overline \gamma_D \to \infty$, which means that the transmit power under OPRA case is close to the average power.

The EC under TCI, where the cutoff level $\gamma_0$ is 0.1, is  smallest except the one under CI in the medium and high SNR region, while this figure for TCI is larger than that under ORA in the low SNR region. It is interesting to note that the EC under CI is zero, and this can be explained by the fact that $\Gamma(j_d) \to \infty$ for $j_d=0$ while $d_0$ is not equal to zero or $K_d$ is not infinity in \eqref{CI}.
\vspace{-1.5em}
\begin{figure}[!htb]
\setlength{\abovecaptionskip}{0pt}
\setlength{\belowcaptionskip}{10pt}
\centering
\includegraphics[width= 3.5 in]{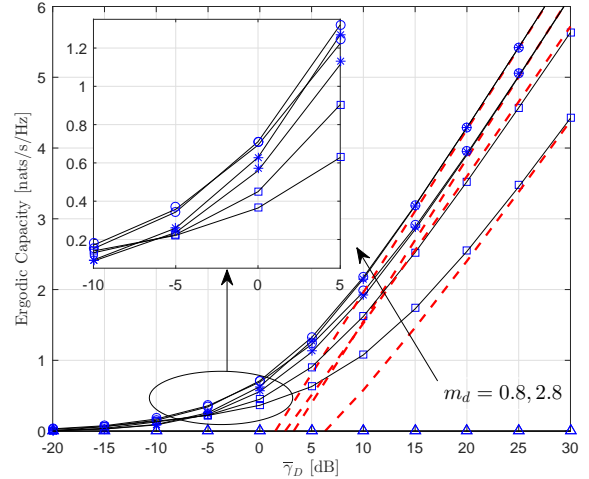}
\caption{Ergodic Capacity versus $\overline \gamma_D$ for $K_d=10$, $\Delta_d=0.5$, $\overline P_t=0$ dB, and $r_d=1$ m, where the circle (star, triangle, and square) symbols, real lines, and dash lines denote simulation, analytical, and asymptotic results under OPRA (ORA, CI, and TCI), respectively.}
\label{fig2}
\end{figure}


 From Fig. 1, our derived asymptotic results  matches very well with the simulation and analytical results in the high SNR region, and the slope of EC under OPRA and ORA is always unity with respect to $\ln \overline \gamma_D$, regardless of parameter settings.


\end{document}